\documentclass[aps,prd,twocolumn,preprintnumbers,superscriptaddress,nofootinbib,floatfix]{revtex4-2}
\usepackage[dvipsnames]{xcolor}
\usepackage{amssymb,amsmath,amsfonts}
\usepackage{bm}
\usepackage{epstopdf}
\usepackage{array}
\usepackage{colortbl}
\usepackage[utf8]{inputenc}
\usepackage{soul}
\usepackage{url}
\usepackage{booktabs}
\usepackage[T1]{fontenc}
\usepackage{longtable}
\usepackage{mathtools}  
\usepackage{fontawesome} 
\usepackage{xspace}
\usepackage{graphicx}
\usepackage{dcolumn}
\usepackage{bm}
\usepackage[colorlinks]{hyperref}
\usepackage{physics}
\usepackage{cleveref}
\usepackage[acronym]{glossaries}

\enlargethispage{\baselineskip}

\makeglossaries
\newacronym{ad}{AD}{automatic differentiation}
\newacronym{bbh}{BBH}{black hole binary}
\newacronym{clt}{CLT}{central limit theorem}
\newacronym{gw}{GW}{gravitational wave}
\newacronym{mc}{MC}{Monte Carlo}
\newacronym{pn}{PN}{post-Newtonian}
\newacronym{snr}{SNR}{signal-to-noise ratio}

\Crefname{equation}{Eq.}{Eqs.}
\Crefname{figure}{Fig.}{Figs.}
\Crefname{section}{Sec.}{Secs.}
\Crefname{tabular}{Tab.}{Tabs.}
\usepackage{siunitx}
\definecolor{pyBlue}{RGB}{31, 119, 180}
\definecolor{pyRed}{RGB}{214, 39, 40}
\definecolor{pyGreen}{RGB}{44, 160, 44}
\definecolor{pyBlue2}{RGB}{0, 111, 237}
\definecolor{pyRed2}{RGB}{224, 52, 36}

\newcommand\myshade{80}
\colorlet{mylinkcolor}{ForestGreen}
\colorlet{mycitecolor}{Red}
\colorlet{myurlcolor}{violet}
\hypersetup{
  linkcolor  = pyGreen!\myshade!black,
  citecolor  = pyBlue!\myshade!black,
  urlcolor   = pyRed!\myshade!black,
  colorlinks = true
}
\newcommand{\neff}{\ensuremath{n_\mathrm{eff}}}

\newcommand{\Ntemplatesrandom}{\ensuremath{{N^\mathcal{R}_T}}}
\newcommand{\Ntemplatesstochastic}{\ensuremath{N^\mathcal{S}_T}}

\def\Msun{{\rm M}_\odot}

\DeclarePairedDelimiter{\ceil}{\lceil}{\rceil}
\newcommand{\jax}{\texttt{jax}\xspace}
\newcommand{\diffbank}{\href{https://github.com/adam-coogan/diffbank}{\texttt{diffbank}}\xspace}

\newcommand{\pintrinsic}{\ensuremath{\bm{\theta}}}
\newcommand{\pextrinsic}{\ensuremath{\bm{\mu}}}
\newcommand{\pall}{\ensuremath{\bm{\Xi}}}
\newcommand{\pcov}{\ensuremath{q}}
\DeclareMathOperator{\mismatch}{m_{\mathrm{mis}}}
\DeclareMathOperator{\targetmismatch}{m_{\mathrm{mis},\ast}}

\begin{document}

\title{Efficient Gravitational Wave Template Bank Generation \\ with Differentiable Waveforms}

\newcommand{\OKC}{\affiliation{The Oskar Klein Centre, Department of Physics, Stockholm University, AlbaNova, SE-106 91 Stockholm, Sweden}} 
\newcommand{\GRAPPA}{\affiliation{Gravitation Astroparticle Physics Amsterdam (GRAPPA), University of Amsterdam, Science Park 904, Amsterdam, 1098 XH, The Netherlands}}
\newcommand{\NORDITA}{\affiliation{Nordic Institute for Theoretical Physics (NORDITA), 106 91 Stockholm, Sweden}}
\newcommand{\UdeM}{\affiliation{Département de Physique, Université de Montréal, 1375 Avenue Thérèse-Lavoie-Roux, Montréal, QC H2V 0B3, Canada}} 
\newcommand{\Mila}{\affiliation{Mila -- Quebec AI Institute, 6666 St-Urbain, \#200, Montreal, QC, H2S 3H1}} 
\newcommand{\CGP}{\affiliation{Center for Gravitational Physics, University of Texas at Austin, Austin, TX 78712, USA}}
\newcommand{\Austin}{\affiliation{Department of Physics, University of Texas at Austin, Austin, TX 78712, USA}}
\newcommand{\IAS}{\affiliation{School of Natural Sciences, Institute for Advanced Study, Princeton, NJ 08540, USA}}
\newcommand{\LIGOMIT}{\affiliation{LIGO Laboratory, Massachusetts Institute of Technology, Cambridge, MA 02139, USA}}

\author{Adam Coogan}\email[Electronic address: ]{adam.coogan@umontreal.ca} \GRAPPA \UdeM \Mila
\author{Thomas D. P. Edwards}\email[Electronic address: ]{thomas.edwards@fysik.su.se} \OKC \NORDITA

\author{Horng Sheng Chia}
\IAS

\author{Richard N.~George}
\CGP

\author{Katherine Freese}
\Austin \OKC \NORDITA

\author{Cody Messick}
\LIGOMIT

\author{Christian N. Setzer}
\OKC

\author{Christoph Weniger}
\GRAPPA

\author{Aaron Zimmerman}
\CGP

\preprint{UTTG 26-2021}
\preprint{NORDITA-2022-004}

\date{\today}

\begin{abstract}
    The most sensitive search pipelines for gravitational waves from compact binary mergers use matched filters to extract signals from the noisy data stream coming from gravitational wave detectors. Matched-filter searches require banks of template waveforms covering the physical parameter space of the binary system. Unfortunately, template bank construction can be a time-consuming task. Here we present a new method for efficiently generating template banks that utilizes automatic differentiation to calculate the parameter space metric. Principally, we demonstrate that automatic differentiation enables accurate computation of the metric for waveforms currently used in search pipelines, whilst being computationally cheap. Additionally, by combining random template placement and a Monte Carlo method for evaluating the fraction of the parameter space that is currently covered, we show that search-ready template banks for frequency-domain waveforms can be rapidly generated. Finally, we argue that differentiable waveforms offer a pathway to accelerating stochastic placement algorithms. We implement all our methods into an easy-to-use Python package based on the \jax framework, \diffbank, to allow the community to easily take advantage of differentiable waveforms for future searches.
\end{abstract}


\maketitle

\section{Introduction}


The detection of \gls*{gw} emission from the binary coalescence of two black holes~\cite{LIGOScientific:2016aoc,LIGOScientific:2021usb,LIGOScientific:2020ibl} opened a new observational window onto the universe. 
To extract \gls*{gw} signals from the noisy data stream, the LIGO and Virgo collaborations typically employ matched filtering.\footnote{
    The matched filter is the optimal linear filter for maximising the signal-to-noise ratio of a known signal in Gaussian-distributed noise~\cite{matched_filter_original}. Note that the matched filter is generally not optimal in the non-Gaussian regime and when searching for multiple possible signals~\cite{Yan:2021wml}.
}
Here, the strain data are compared to a bank of templates described by a set of points in the binary parameter space and a \gls*{gw} waveform model. The goal of template bank generation is to have at least one sufficiently similar template in the bank for any potential signal \cite{PhysRevD.44.3819}. 
The challenge is to do this with a minimal number of templates and computing resources.
To date, matched-filter searches for transient \gls*{gw} emission from compact binary coalescences have focused on aligned-spin \gls*{bbh} systems on quasi-circular orbits.\footnote{
    Binary black hole templates are sufficiently similar to those from binary neutron star and black hole - neutron star inspirals such that dedicated template banks are not necessary~\cite{Harry:2018hke,Chia:2020psj}.
} Significant work has therefore been put into carefully constructing close-to-optimal template banks for the \gls*{bbh} parameter space. 
Conversely, less work has been done to carry out searches from more general systems such as BBHs with generic spins (see e.g.~\cite{gr-qc/0211087,1411.6815,1612.05173,1603.02444}) or binaries that contain objects which significantly differ from black holes such as boson stars or black holes with superradiant clouds~\cite{Chia:2020psj}.
The focus of this paper is to take a first step towards enabling fundamentally new searches, including both Beyond the Standard Model physics and new astrophysics, by making the generation of template banks for new systems efficient and simple. 

Constructing template banks for realistic \gls*{gw} waveforms is a challenging task with many different methods currently being employed. 
It is particularly difficult since the goal of most searches is to minimize the computational cost while achieving a given detection efficiency.
Having a small number of templates is both statistically preferable (reduces the number of trials) and computationally preferable (computing the matched filter repeatedly is expensive). 
Template placement can be therefore thought of as a form of the mathematical sphere-covering problem on a parameter space manifold with distances given by the \emph{match}~\cite{Owen:1995tm}, which measures the similarity between different gravitational waveforms and is reviewed in \cref{sec:metric}.

If the parameter space manifold is sufficiently flat, optimum solutions exist for low numbers of dimensions~\cite{Prix:2007ks} and can be realized through lattice placement algorithms which utilize a parameter space metric to guide the template placement~\cite{Cokelaer:2007kx}. On the other hand, for spaces that are curved or have complicated boundaries, lattice-based template banks are difficult to construct~\cite{Manca:2009xw}. 

An alternative is stochastic placement~\cite{Harry:2009ea}.\footnote{
    More recently hybrid placement schemes have been used in the literature~\cite{Roy:2017oul,Roy:2017qgg, treebank} along with the development of a geometric placement algorithm~\cite{Roulet:2019hzy}.
} Stochastic placement works by randomly proposing template positions in the parameter space which are only accepted if they are sufficiently far from all other templates in the bank. The requirement of complete coverage of the parameter space is abandoned in favor of requiring coverage with a chosen level of confidence. Although stochastic banks are in principle simple to construct and do not require knowledge of the parameter space's geometry, they become computationally expensive when the number of templates becomes large due to the large number of match calculations needed to check whether a template should be accepted or rejected. Moreover, if the trial waveforms are sampled uniformly from the parameter space, the acceptance rate may be very small, further slowing convergence. This acceptance rate can be improved if one uses the metric density as a probability distribution on the space of parameters~\cite{Messenger:2008ta}. 

We have encountered practical difficulties in our own work constructing template banks for new searches in LIGO data, particularly when searching for compact objects with large spin-induced quadrupole moments~\cite{waveformpaper,searchpaper}.\footnote{
    This might be due to our aim to search for spin-induced quadrupole numbers up to $\kappa\sim10^3$.
} While the computational cost of performing matched filtering far exceeds the cost of generating a stochastic template bank, matched filtering is straightforward to parallelize while template bank generation is not.\footnote{
    One option for parallelizing bank generation is to partition the parameter space and generate banks for each of those subspaces. This can substantially decrease bank generation time at the cost of increasing the size of the bank. However, coming up with an effective partition in general is complicated.
} This means that the wall time\footnote{
    The ``wall time'' for a computer program is the time elapsed from the beginning to end of the program's execution. For parallel programs, this is different from ``CPU time'', which is the amount of execution time for each CPU used for the program.
} to generate stochastic banks can make up a nontrivial fraction of the time required to conduct a search. We have found this to be the case when employing existing bank generation code used by the LIGO, Virgo and KAGRA collaborations\footnote{
    See \href{https://github.com/lscsoft/lalsuite/blob/master/lalapps/src/inspiral/lalapps_cbc_sbank.py}{lalapps\_cbc\_sbank.py} in LALSuite.
} in our own work conducting searches for new physics in LIGO data. 
Meanwhile, we found it difficult to integrate the efficient geometric template bank generation scheme from Ref.~\cite{Roulet:2019hzy} into existing \gls*{gw} search pipelines (in particular \texttt{GstLal}), as the method generates templates in a lower-dimensional space that cannot easily be mapped back to physical binary parameters.

Motivated by these issues, in this paper we instead explore the random bank generation method, first described in Ref.~\cite{Messenger:2008ta}. We build upon this method in two key ways. First, we use introduce \emph{automatically-differentiable waveforms}. These enable us to automatically compute the parameter space metric and therefore the sampling probability distribution. Second, we introduce a new stopping criterion that naturally accounts for parameter space boundaries without the need to fine tune the bank generation method for each waveform model and parameter space. Together these improvements make random bank generation simple and efficient for both \gls*{bbh} \gls*{gw} models as well as more exotic scenarios~\cite{Chia:2020psj}.

\Gls*{ad} is a foundational computational tool in modern machine learning~\cite{DBLP:journals/corr/BaydinPR15}. \Gls*{ad} is an approach to exactly calculating\footnote{
    Exact up to floating point errors.
} derivatives of functions defined by computer programs. It leverages the fact that any program can be broken into elementary operations which are each differentiable and whose derivatives can be combined using the chain rule. This enables differentiation for $\order{1}$ additional computational cost beyond the function evaluation itself. It is distinct from numerical differentiation (which is based on finite differencing, causing it to be numerically unstable and scale poorly with dimensionality\footnote{
    More precisely, finite differencing requires at least two function evaluations per parameter.
}) and symbolic differentiation (which can yield cumbersome expressions that must be hand-coded). In machine learning, \gls*{ad} libraries~\cite{jax2018github,pytorch,zygote,forwarddiff,enzyme} enable the training of large neural networks via gradient descent. More broadly, the nascent field of \emph{differentiable programming} combines \gls*{ad} with programs beyond neural networks, such as physics simulators (see e.g.~\cite{Hu2019-xc}). Differentiable programming has only been applied to a few domains within astrophysics so far (see e.g.~\cite{Chianese:2019ifk,Coogan:2020yux,Karchev:2021fro,eosdp,Bohm:2020ilt,2021PASP..133c4505M,Hearin:2021tsy,2021ApJ...907...40P,AlvesBatista:2021gzc,AlvesBatista:2021gzc}), where it enables fast, fully-automated fitting and approximate Bayesian inference. Here, we utilize the Python \gls*{ad} framework \jax~\cite{jax2018github} to create differentiable frequency-domain waveforms so that we can automatically compute their parameter space metric. We demonstrate our metric calculation is extremely accurate and fast.

To naturally account for parameter space boundaries we introduce a new \gls*{mc} method to track the fraction of the parameter space covered as the bank is generated. Instead of precomputing the number of templates required to cover the parameter space, as is done in Ref.~\cite{Messenger:2008ta}, we start with a number of \emph{effectualness points} which are iteratively removed once they are covered by a template. Although these random banks contain more templates than stochastic ones, we show that their generation can be be orders of magnitude faster. Moreover, we argue that by using the metric, the time taken to construct a bank stochastically can also be reduced.

Finally, to help the community utilize our methods, we present our bank generation code in an easy-to-use Python package called \diffbank~\cite{diffbank}.

The remainder of this paper is structured as follows. In \Cref{sec:metric} we define the metric and discuss the accuracy, speed, and limitations of \gls*{ad}. In \Cref{sec:randombanks} we present our ``effectualness points'' approach to template bank construction and discuss its scaling behaviour. We compare banks generated with our method to others in the literature for waveform models of different dimensionalities in \Cref{sec:comparison}. Finally, we conclude in \Cref{sec:conclude}.

\section{Parameter Space Metric and Automatic Differentiation}
\label{sec:metric}

In this section we review the definition of the parameter space metric, discuss the accuracy of the \gls*{ad} metric, its computational speed, and the limits of \gls*{ad}.

\subsection{Defining the Parameter Space Metric}

Throughout this work we focus on frequency domain waveforms which can be expressed in the form
\begin{equation}
  h_{\pall}(f) = A_{\pall}(f) e^{i \Psi_{\pall}(f)} \, , \label{eqn:freq_wvf}
\end{equation}
where $A_{\pall}(f)$ is the amplitude of the waveform, $\Psi_{\pall}(f)$ is the phase and the subscript indicates dependence on a set of parameters $\pall$. To allow a comparison between two frequency domain waveforms, $h_{\pall_1}(f)$ and $h_{\pall_2}(f)$, one begins by defining the noise weighted inner product
\begin{equation}
  \left(h_{\pall_1}|h_{\pall_2}\right) \equiv 4 \, \mathrm{Re} \int^{\infty}_{0} \mathrm{d} f \, \frac{ h_{\pall_1}(f) h^*_{\pall_2}(f)}{S_n(f)}\, ,
  \label{eqn:innerprod}
\end{equation}
where $S_n$ is the (one-sided) noise power spectral density (PSD). We can normalize the inner product through
\begin{equation}
    \left[h_{\pall_1}|h_{\pall_2}\right] \equiv (\hat{h}_{\pall_1} | \hat{h}_{\pall_2} ) = \frac{\left(h_{\pall_1}|h_{\pall_2}\right)}{\sqrt{\left(h_{\pall_1}|h_{\pall_1}\right)\left(h_{\pall_2}|h_{\pall_2}\right)}}\, , \label{eqn:dotprodnorm}
\end{equation}
where we introduced the normalized waveform $\hat{h}_{\pall} \equiv h / (h_{\pall} | h_{\pall})^{1/2}$.
For binaries on circular orbits observed by a single detector, the waveform parameters can be split into a set of intrinsic parameters $ \pintrinsic$ (i.e.~properties of the binary system such as masses and spins) and two extrinsic parameters $\pextrinsic = (t_{c}, \phi_{c})$\footnote{
    It is common in the \gls*{gw} community to say there are seven extrinsic variables: $\{D, \alpha, \delta, \psi, \iota, \phi, t_c\}$. Respectively, these are the luminosity distance, right ascension, declination, polarisation angle,  inclination angle, phase at some reference time, and time of coalescence.   Since $\{D, \alpha, \delta, \psi, \iota, \phi\}$ only affect the waveform through an overall phase and amplitude constant (under the assumption that the detector's response is constant over the duration of the waveform), these parameters can be absorbed into a combination of $\phi_c$ and an amplitude normalization.
}, the time and phase at coalescence for the waveforms. A quantity commonly used to characterize the difference between waveforms with different intrinsic parameters is the \emph{match}~\cite{Owen:1995tm}, given by maximizing the inner product over the extrinsic parameters:
\begin{align}
    \operatorname{m}(\pintrinsic_1 , \pintrinsic_2) &\equiv \max_{\pextrinsic_1,\, \pextrinsic_2} \left[h_{\pall_1} \, \middle| \, h_{\pall_2} \right] \\
    &= \max_{\Delta t_{c},\, \Delta\phi_{c}} \left[ h_{\pintrinsic_1, \pextrinsic_1 = 0} \, \middle| \, h_{\pintrinsic_2, \pextrinsic_2 = 0} \ e^{i (2 \pi f \Delta t_c + \Delta \phi_c)} \right] \, , \label{eqn:match}
\end{align}
where $\Delta t_c = t_{c,1} - t_{c,2}$ is difference between the time of coalescence for the two waveforms (similarly for the phase). The second equality comes from the fact that the extrinsic parameters enter only in a factor $\exp[i (2 \pi\, f\, \Delta t_c + \Delta \phi_c)]$ in the inner product.\footnote{This assumes a $(\ell |m|) = (22)$-only waveform.} Since $\phi_c$ only appears in the inner product in the overall phase and the inner product integrand takes the form of a Fourier transform, the maximization over $\phi_c$ can be efficiently accomplished by taking the absolute value of the Fourier transform of the normalized inner product~\cite{Schutz_GW}~(see also Ref.~\cite[p.~388]{Maggiore:2007ulw}). The \emph{mismatch distance} between waveforms can then be simply defined as $\mismatch \equiv 1 - m$. 

For concreteness we will use subscript 1 to refer to the observed signal $h_{\pall_1}$ and subscript 2 to refer to waveforms from the template bank $h_{\pall_2}$. To quantify how well a template bank can recover signals, we introduce the \emph{effectualness} $\varepsilon$~\cite{Damour:1997ub}, which involves an additional maximization over the intrinsic parameters of the waveforms in the bank:  \begin{equation}
    \varepsilon(\pintrinsic_1) \equiv \max_{ \pextrinsic_1,\, \pall_2} \left[h_{\pall_1} \, \middle| \, h_{\pall_2} \right] = \max_{\pintrinsic_2} \operatorname{m}(\pintrinsic_1 , \pintrinsic_2) \,
    \label{eqn:effectualness}
\end{equation}
where $\pall \equiv (\pextrinsic, \pintrinsic)$. In general, there is no trick to maximizing over the intrinsic parameters of the bank: a given bank must be constructed and compared with $h_{\pall_1}$. The maximization in the effectualness thus amounts to finding the template in the bank that most closely resembles the signal $h_{\pall_1}$. In other words, the effectualness quantifies the fraction of \gls*{snr} retained when using a discretized template bank to search for a binary system with intrinsic parameters $\pintrinsic_1$. Typically, one wants to ensure that $\varepsilon$ remains above a threshold value throughout $\pintrinsic$. 

In order to construct a template bank, we define the parameter space \emph{metric} $g_{ij}$ over intrinsic parameters~\cite{Owen:1998dk,Cokelaer:2007kx,Manca:2009xw,Keppel:2013kia,Messenger:2008ta,Hanna:2021luk}. We start by noting that for small differences in the intrinsic parameters we can Taylor-expand the mismatch distance to quadratic order as
\begin{equation}
    \mismatch(\pintrinsic, \pintrinsic + \Delta\pintrinsic) \equiv 1 - \operatorname{m}(\pintrinsic, \pintrinsic + \Delta\pintrinsic) \approx g_{ij}(\pintrinsic) \, \Delta\theta^i \, \Delta\theta^j \, . \label{eq:mismatch-distance}
\end{equation}
The metric is thus related to the Hessian of the match:
\begin{equation}
    g_{ij}(\pintrinsic) \equiv -\frac{1}{2} \eval{\pdv{\operatorname{m}(\pintrinsic, \pintrinsic + \Delta\pintrinsic)}{\Delta\theta^i}{\Delta\theta^j}}_{\Delta\pintrinsic = 0} \, .
\end{equation}
We use Latin letters to index the intrinsic parameters in the full parameter vector $\pall$. Hence, for a given functional form of the waveform in \Cref{eqn:freq_wvf}, we can construct a match between two neighboring points and then from its derivatives, obtain the metric in parameter space.

We can simplify our calculation of the metric $g_{ij}$ in the following way. The match in \Cref{eqn:match} is obtained by maximization over $\Delta \phi_c$ and $\Delta t_c$. As explained above, analytically maximizing the match with respect to $\Delta \phi_c$ is straightforward. Hence we construct a new metric $\gamma_{IJ}(\Delta t_c, \pintrinsic)$ that is a function of $\Delta t_c$ in addition to the instrinsic parameters. We use $\bm{\Theta} \equiv (\Delta t_c, \pintrinsic)$ to denote the concatenation of the difference in coalescence times with the set of intrinsic parameters and index $\bm{\Theta}$ with capital Latin indices, with the first index being zero and the others running over the intrinsic parameters. With this notation, the metric over $\bm{\Theta}$ is given by
\begin{align}
\gamma_{IJ}(\bm{\Theta}) &\equiv -\frac{1}{2} \eval{\pdv{ \max_{\Delta \phi_c} \left[ h_{\pall_1} \middle| h_{\pall_2} \right] }{\Delta\Theta^I}{\Delta\Theta^J}}_{\bm{\Theta} = 0} \\
    &= -\frac{1}{2} \eval{\pdv{ \left| \left[ h_{\pall_1} \middle| h_{\pall_2} \right] \right| }{\Delta\Theta^I}{\Delta\Theta^J}}_{\bm{\Theta} = 0} \, ,
\end{align}
where the absolute value accomplishes the maximization over $\Delta\phi_c$. 
The intrinsic metric is then the projection of this metric onto the subspace orthogonal to $\Delta t_c$~\cite{Owen:1995tm,Carroll:2004st}:
\begin{equation}
    g_{ij} \equiv \gamma_{ij} - \frac{\gamma_{0 i} \gamma_{0 j}}{\gamma_{00}} \, ,
\end{equation}
where the index $0$ corresponds to $\Delta t_c$.\footnote{
    Alternatively, as discussed above we could instead compute the metric by maximizing the match over $\Delta t_c$ with a Fourier transform and differentiating through this operation. While both methods give metrics that agree to within a few percent for the waveforms we tested, we found differentiating through the Fourier transform to be slower than analytically maximizing over $\Delta t_c$.
}
Computation of the metric $g_{ij}$ via Eqn (10) is more efficient (and numerically more accurate) than directly computing Eqn. (8).

Finally, we define the absolute value of the metric determinant $g \equiv |\det g_{ij}|$, the square root of which describes the scaling of a volume element between match space and the parameter space. More concretely, the proper volume $V$ of the parameter space $S$ is given by
\begin{equation}
    V_S = \int_S \dd{V}, \quad \text{with} \quad \dd{V} \equiv \sqrt{g}\, \dd[n]{\pintrinsic} \,.
\end{equation} 
Sampling from the probability density function generated by the metric determinant across the parameter space (i.e.~the \emph{metric density}) therefore corresponds to uniform sampling in proper volume. 

\subsection{Computing the Parameter Space Metric with Automatic Differentiation}
\label{sec:limAD}

We can use the metric in a variety of ways to construct a template bank. For example, in parameter spaces that are sufficiently flat, lattice placement algorithms can be used to optimally place templates. Unfortunately, parameter space boundaries often complicate these placement schemes. The most common schemes used today are hybrid methods which start out with a lattice scaffold which are then supplemented using a stochastic placement algorithm~\cite{Roy:2017oul,Roy:2017qgg}. Though, as mentioned above, stochastic placement schemes are slow and can be even slower when the sampling distribution is very different to the target distribution.

For waveforms which can be written in closed form (e.g. \gls*{pn} waveforms covering the inspiral phase), the metric can be decomposed in terms of a finite number of integrals that can be numerically precomputed. While this is practical for simple models, closed-form models continue to grow in complexity (see e.g.~\cite{2004.06503}), making this approach unsustainable. On the other hand, for complex waveform models one typically must compute the metric through numerical differentiation~\cite{Roy:2017oul}, which is numerically unstable and scales poorly with dimensionality. We propose to replace these bespoke approaches with automatic differentiation (AD). 

\gls*{ad} is a modern computational framework which has been extensively used throughout the machine learning community due to its ability to differentiate through arbitrarily complex functions at little extra cost over evaluating the function itself. Its popularity has now led to the development of several efficient implementations in both Python and Julia programming languages. We choose to implement our metric calculation using \jax~\cite{jax2018github} due to its native integration with Python, its speed, and its native GPU support. We discuss the benefits and limitations of various \gls*{ad} implementations below. 

\begin{figure}
    \includegraphics[width=0.97\linewidth]{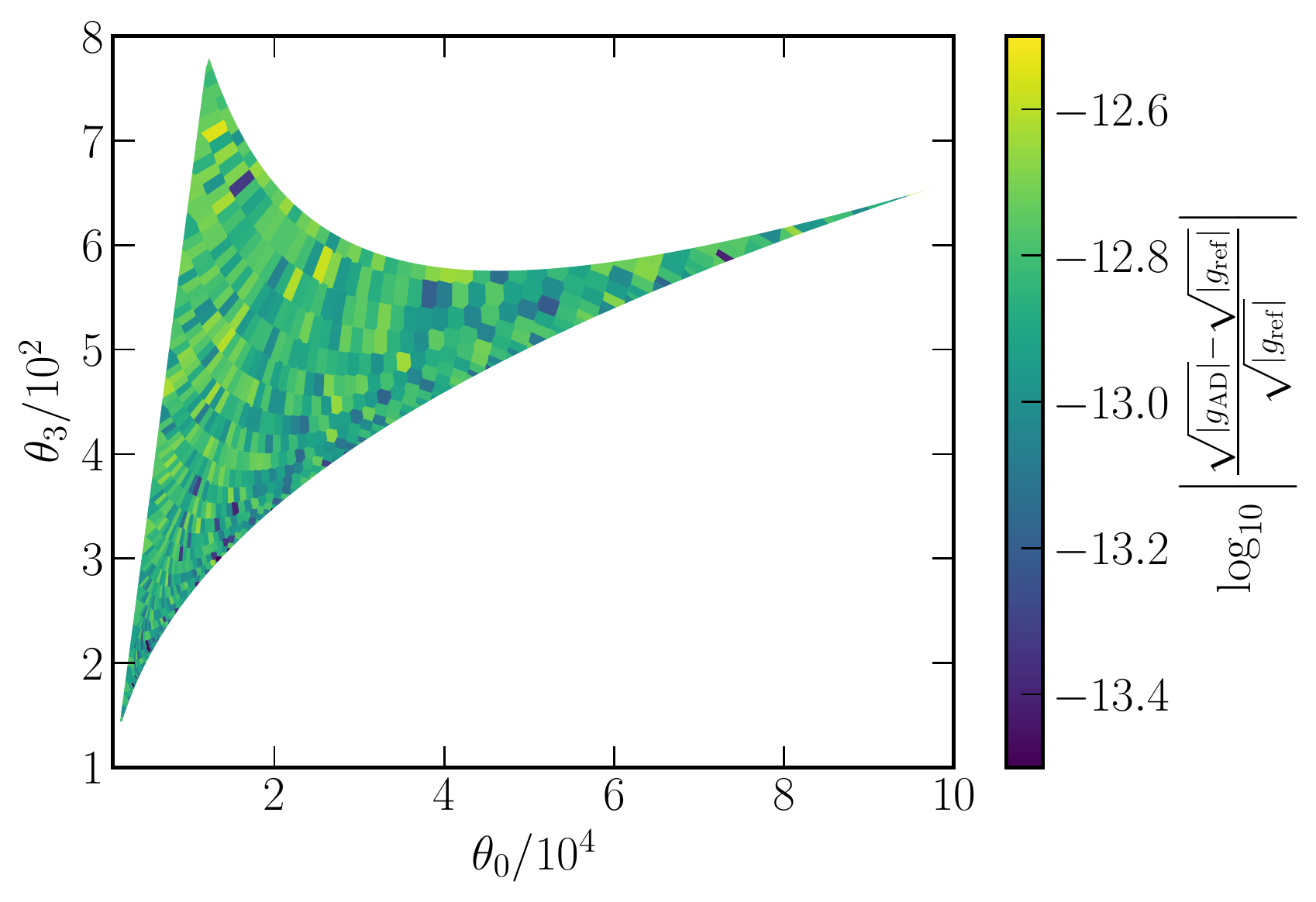}
    \caption{Comparison of the metric computed using automatic differentiation with an analytic reference metric~\cite{Babak:2006ty}, parametrized in terms of dimensionless chirp times~\cite{Owen:1998dk}. Both were computed using 64-bit precision floating point numbers, the precision of which is about 15 significant digits. The noise model is described in the text, and the frequency grid spans \SIrange{10}{512}{\hertz} with a spacing of \SI{0.1}{\hertz}. The colored regions are a Voronoi tessellation. \href{https://github.com/adam-coogan/diffbank/blob/paper-1/scripts/plot_density.py}{\faFileCodeO}}
    \label{fig:density}
\end{figure}

As an illustration of the accuracy of the \gls*{ad}-computed metric, we compare to the analytically computed metric determinant for the 2PN frequency-domain waveform used to describe the inspiral of nonspinning quasicircular binary black holes~\cite{Babak:2006ty}. This metric was derived in the dimensionless chirp time coordinates ($\theta_0$ and $\theta_3$)~\cite{Owen:1998dk,Mohanty:1997eu} which were hand-chosen to keep the metric as flat as possible.\footnote{
    Note that Ref.~\cite{Babak:2006ty} uses the notation $\theta_1$ and $\theta_2$, which we instead refer to as $\theta_0$ and $\theta_3$. This choice was made to match onto the subscript notation used for $\tau_0$ and $\tau_3$ -- see Eq.~(3.18) of Ref.~\cite{Babak:2006ty}.
}
In \Cref{fig:density}, we show the relative error between the \gls*{ad} computed metric determinant and the analytic calculation from Ref.~\cite{Babak:2006ty}, where we have used a PSD representative of the Livingston detector during O3a.\footnote{
    \url{https://dcc.ligo.org/LIGO-P2000251/public}
} The relative differences are extremely small and, for much of the parameter space, only a few digits above the precision of 64-bit floating point numbers. We performed similar checks for a variety of other waveforms, finding similar results for all.

As mentioned above, a key feature of \gls*{ad} is the speed of evaluation. For this particular waveform we find that a single metric evaluation takes approximately \SI{e-3}{\second} on a Intel Xeon CPU E5-2695 v4 with a 2.10\,GHz clock speed or an NVidia V100SXM2 GPU. 
This will of course increase with increasingly complex waveforms, but is easily efficient enough for our purposes and, as we show, can be used to produce template banks extremely efficiently. 

\begin{figure}
    \includegraphics[width=0.97\linewidth]{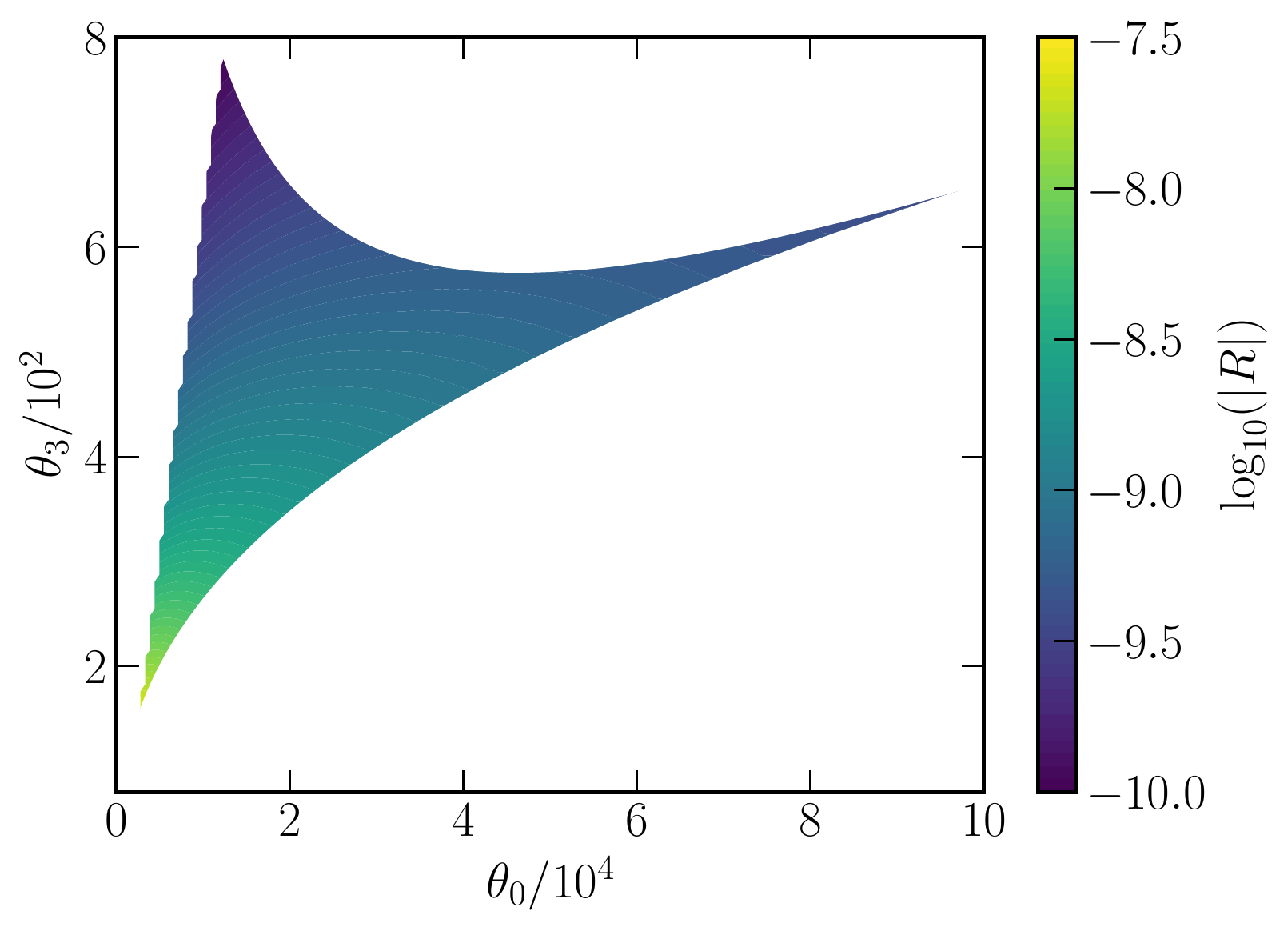}
    \caption{The scalar curvature for the waveform from Ref.~\cite{Babak:2006ty}, parametrized in terms of dimensionless chirp times~\cite{Owen:1998dk}. The configuration for the metric calculation is described in the caption for \Cref{fig:density}. \href{https://github.com/adam-coogan/diffbank/blob/paper-1/scripts/plot_scalar_curvature.py}{\faFileCodeO}}
    \label{fig:scalar-curvature}
\end{figure}

This speed also gives us easy access to other geometric quantities that measure how non-Euclidean the parameter space manifold is. For instance, in \Cref{fig:scalar-curvature} we plot the scalar curvature for the same 2PN waveform as is shown in \Cref{fig:density}. The scalar curvature is a simple coordinate-invariant quantity that encodes the local geometry of the manifold and could be used to determine how non-Euclidean a parameter space is. We provide a small Python package, \href{https://github.com/adam-coogan/diffjeom}{\texttt{diffjeom}}~\cite{diffjeom}, to facilitate calculating such quantities.

\vskip 3pt

In principle, there are no limitations to what can be differentiated with \gls*{ad}, since any program can be decomposed into a set of fundamental operations which are each differentiable.\footnote{
    Of course, some functions (such as $f(x) = |x|$) are non-differentiable. This does not matter in practice, however, since it is extremely unlikely that a function will need to be evaluated at a point where it is non-differentiable.
} In practice, different \gls*{ad} implementations such as \texttt{pytorch}~\cite{pytorch}, \jax~\cite{jax2018github}, \texttt{zygote}~\cite{zygote}, \texttt{ForwardDiff}~\cite{forwarddiff} and \texttt{enzyme}~\cite{enzyme} have their own restrictions.

In this work we use \jax due to its \texttt{numpy}-like interface and ability to just-in-time-compile code to run very efficiently. At minimum, the metric for any waveform that can be expressed in closed form can be computed using \gls*{ad}. \jax currently has limited support for special functions, which can complicate the implementation of waveforms involving them. A practical (but not fundamental) roadblock in implementing special functions is that \jax's compiler only supports a limited type of recursion called tail recursion. While all recursive programs can in principle be expressed using tail recursion, performing the conversion manually can be labor-intensive and lead to excessive amounts of code.

The most complicated waveforms such as the effective one-body (EOB) formalism~\cite{Buonanno:1998gg} involve difficulties aside from the use of special functions. EOB waveforms are generated by solving a set of differential equations (e.g.~\cite{2004.09442,2101.08624}) involving complex numbers at intermediate stages of the calculation, special functions, and root-finding to set the initial conditions. 
In principle, \jax and other frameworks are capable of differentiating through these operations. In this work, we focus on closed-form PN frequency domain waveforms and leave the task of making more complex differentiable waveforms for future investigation.

\section{Random Bank Generation}
\label{sec:randombanks}

In this section we describe our bank generation procedure and discuss the general features of the banks it produces.

\begin{figure*}
    \includegraphics[width=0.99\textwidth]{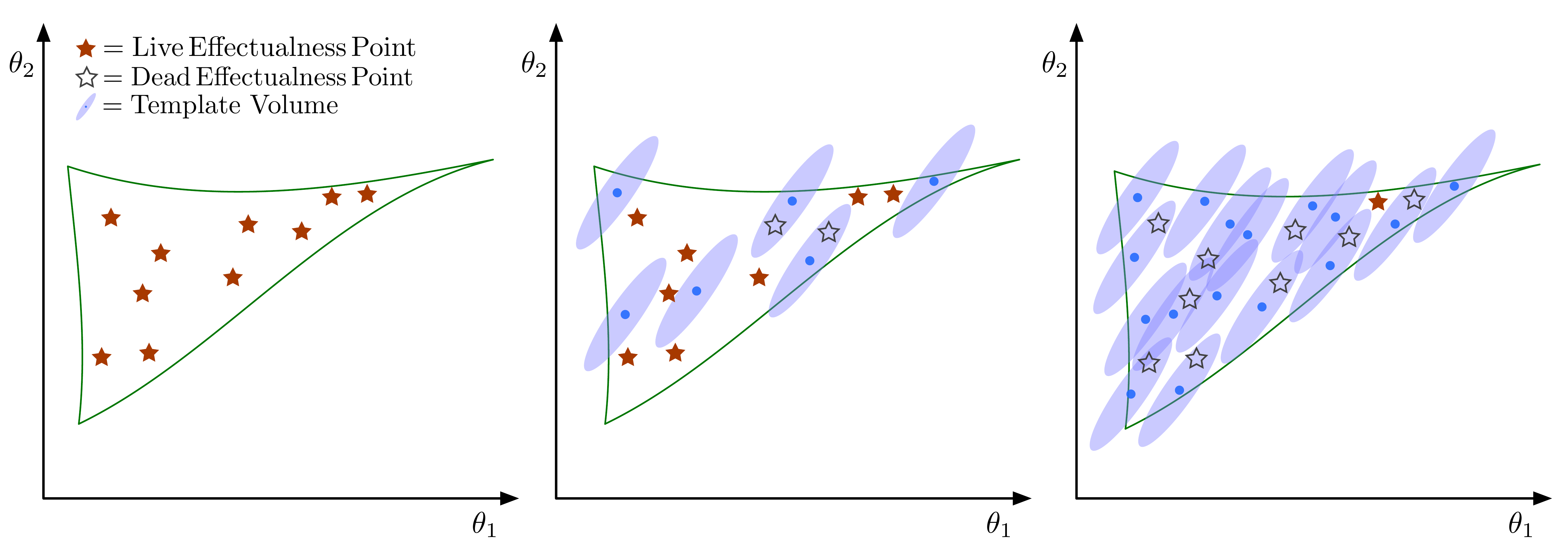}
    \caption{A schematic illustration of our template bank generation procedure. The bank generation starts with $\neff$ live effectualness points which are sampled according to the metric density $p(\pintrinsic) \propto \sqrt{g(\pintrinsic)}$ (illustrated by the filled red stars in the left panel). The solid (green) curves in the shape of a distorted triangle illustrate the boundary of the parameter space.  Templates (illustrated by the blue points with accompanying purple ellipses with radial scale $\sqrt{\targetmismatch}$) are then added sequentially where the template's position is also sampled from the metric density. Each time a template is added, the match is calculated for all live effectualness points. If the match is greater than $1-\targetmismatch$ for some effectualness point, then that point is indeed successfully described by the new template;  that effectualness point  ``dies'' (illustrated by the empty stars in the central panel which are covered by at least one template). This sampling continues until $\ceil{\eta \, \neff}$ effectualness points are covered (illustrated in the right panel for $\eta=0.9$ and $\neff=10$). Hence the choice of $\eta$ determines our stopping criterion. Note here we have taken the orientation and size of the template ellipses to be constant. In general both will vary over the parameter space, further complicating bank generation.}
    \label{fig:eff_points_illustration}
\end{figure*}

\subsection{Generating a Bank}
\label{sec:generating-a-bank}

As mentioned above, our method builds upon that of Ref.~\cite{Messenger:2008ta}, which randomly samples a predetermined number of templates to achieve a target coverage probability. Interestingly, Ref.~\cite{Messenger:2008ta} also shows that in high dimensions, at the expense of not covering the entire parameter space, these random template banks can actually contain significantly fewer templates than similar banks made using optimal lattice placement algorithms while achieving slightly less than full coverage.

Two user-determined quantities control the generation of a random template bank: a covering fraction $\eta$ and the target maximum mismatch $\targetmismatch$, where the mismatch is defined in \Cref{eq:mismatch-distance}. For a random bank, a point in parameter space is {\it covered} by a template if the mismatch $\mismatch$ between the point and template obeys $\mismatch < \targetmismatch$. Equivalently, the point is covered if it lies within the $n$-dimensional ellipsoid of scale $\sqrt{\targetmismatch}$ centered on the template.\footnote{
    Note that technically the ellipsoid is defined by an $n$-dimensional hypersphere with radius $\sqrt{\targetmismatch}$ which is transformed into an ellipsoid using the metric. Equivalently, the axes of the ellipsoid can be defined by scaling the eigenvectors of the metric by $\sqrt{\targetmismatch / \lambda_i}$, where $\{ \lambda_i \}$ are the eigenvalues of the metric.
}
Hence, the problem of template bank generation relates to the sphere-covering problem, which seeks to find the smallest number of spheres to cover an $n$-dimensional Euclidean space.\footnote{
    The metrics in this paper are typically curved rather than flat. However, in cases where the metric is flat, one can use the metric to transform to Euclidean space so that the covering problem at hand is indeed an example of the sphere covering problem in Euclidean space. Note, however, that optimal lattice placement schemes can still be tricky to implement when the boundaries of the parameter space are non-trivial.
}
However, random banks do not attempt to cover the whole parameter space. Instead, the probability that a given point in parameter space is covered by a template is approximately $\eta$.

Reference~\cite{Messenger:2008ta} found that by sampling template positions from the metric density, there is a direct relationship between the number of templates sampled, the covering fraction $\eta$, and the total volume of the space. One can therefore precompute the number of templates necessary to achieve a target $\eta$. However, the analysis in Ref.~\cite{Messenger:2008ta} assumes that a template's volume (the interior of the ellipse defined by the eigenvectors of the metric and scaled by $\sqrt{\targetmismatch}$; see \Cref{fig:eff_points_illustration} for an illustration) is entirely contained within the boundaries of the parameter space. Unfortunately, for nearly all standard parameter spaces considered in \gls*{gw} physics, this assumption does not hold. This motivates us to define a new stopping criterion which accounts for the fraction of a template's volume remaining within the boundary.

Our procedure aims to generate a random bank with predetermined target values of the covering fraction $\eta$ and maximum mismatch $\targetmismatch$. From a high level, we start by sampling a set of points within the parameter space which we then compare to templates as they are sequentially added to the bank. The goal is to construct a running \gls*{mc} estimate for the fraction of the volume of parameter space covered by at least one of the templates in the bank, and to stop when the predetermined target value $\eta$ is reached.  

In more detail, we start by generating a set of $\neff$ \emph{effectualness points}, with positions randomly sampled according to the metric density $p(\pintrinsic) \propto \sqrt{g(\pintrinsic)}$.
To generate the bank we sequentially add new templates, again with positions randomly sampled according to the metric density.
For each new template, we check whether or not it covers any of the effectualness points.  
As previously mentioned, each template has an accompanying ellipsoid with its radial scale set by $\sqrt{\targetmismatch}$.
Any effectualness points covered by the template's ellipsoid are removed from the generation process and no longer compared to newly added templates (i.e.~they are dead).
This process is repeated until the fraction of dead effectualness points divided by the total initial number of effectualness points $\neff$ is less than the target value $\eta$.  In other words, we reach our stopping criterion when $\ceil{\eta \, \neff}$ points are covered. 

We schematically illustrate the bank generation procedure in \Cref{fig:eff_points_illustration} for a two-dimensional parameter space. Here, the filled red stars indicate live effectualness points and the empty grey stars indicate dead ones. The distorted green triangle outlines the boundary of the parameter space. The blue points (with accompanying purple ellipsoids) represent templates and their accompanying volume, the scale of which is set by $\sqrt{\targetmismatch}$.

Each panel represents a different stage in the bank generation procedure. The left panel shows the initialization with only live effectualness points. The middle panel represents an intermediate stage where some effectualness points are dead since they have been covered by newly-added templates. The right panel shows the bank once the stopping criterion has been met for a target $\eta = 0.9$ (i.e., 90\% of the effectualness points are covered). The bank generation relies solely on computing the number of covered effectualness points when we add templates. The remaining white region at the end of the bank generation is the portion of parameter space that remains uncovered and contains $(1 - \eta) \, \neff$ of the original points.

\begin{figure*}
    \centering
    \includegraphics[width=0.8\linewidth]{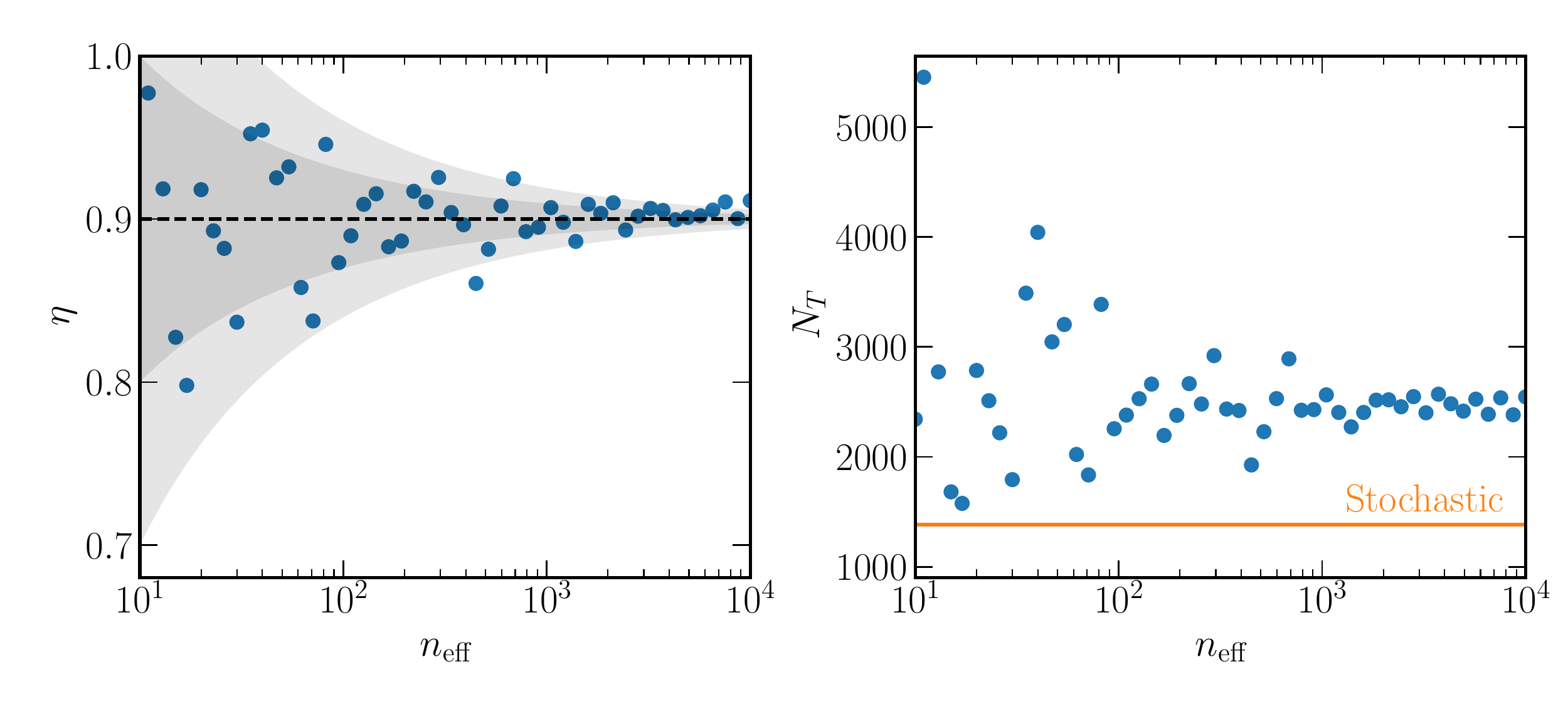}
    \caption{\emph{Left panel:} Covering fraction $\eta$ of random banks as a function of the number of effectualness points $\neff$ using the model configuration explained in \Cref{sec:scaling}. The blue points indicate estimates of $\hat{\eta}$ calculated using an additional set of randomly injected points (i.e., not the effectualness points used to generate the bank). The grey bands show the one and two $\sigma_{\hat{\eta}}$ error bands on $\hat{\eta}$. Note that for $\neff \sim \mathcal{O}(10)$, the error extends to the unphysical region $\eta \geq 1$. This could be corrected by using an asymmetric estimate of $\sigma_{\hat{\eta}}$ such as the Jeffreys interval~\cite{JeffreysInterval}). \emph{Right panel:} The number of templates in the bank required to meet the stopping criterion as a function of $\neff$. For comparison, we also plot the number of templates in a stochastic bank (orange line) which was generated with $\eta=0.9$. Note that in this example we find that $\Ntemplatesrandom > \neff$ for $\neff\sim\mathcal{O}(10^4)$. In this case, the scaling estimates discussed in \Cref{sec:scaling} will be inaccurate since a single template is likely to cover multiple effectualness points. \href{https://github.com/adam-coogan/diffbank/blob/paper-1/scripts/plot_neff_scaling.py}{\faFileCodeO}}
    \label{fig:eta-dist}
\end{figure*}

At the start of bank generation, one is required to perform a match calculation for each effectualness point for every template added to the bank. 
Fortunately, this is offset by the fact that the probability of a template covering an effectualness point is initially large, falling only as effectualness points are removed. Since our generated banks are based on an \gls*{mc} estimate of $\eta$, the randomness of the stopping criteria leads to a realized covering fraction $\hat{\eta}\sim\eta$, with the associated \gls*{mc} error derived in \cref{sec:eta-error}
\begin{equation}
    \label{eq:sigma-eta-hat}
    \sigma_{\hat{\eta}} \approx \sqrt{\frac{(1 - \eta) \eta}{\neff - 1}} \, ,
\end{equation}
where we assume the errors are Gaussian. Importantly, since this is an \gls*{mc} estimate, it does not depend on the dimensionality or volume of the parameter space. Since this error tends to zero as $\neff$ approaches infinity, using a large number of effectualness points yields a bank with coverage fraction near the target $\eta$. One therefore needs to choose an $\neff$ large enough that $\sigma_{\hat{\eta}}$ is small, but small enough that the time to generate the bank is not too long. This effect is clearly illustrated in \Cref{fig:eta-dist}, for which we generate a number of random banks with different $\neff$ and the same target coverage $\eta = 0.9$, together with a reference stochastic bank. 

For each bank we use a separate set of injected points to measure the achieved $\hat{\eta}$ (shown as blue dots in the left hand panel). Each random bank produces a different realization of $\hat{\eta}$ and a different number of templates. We see however, that $\hat{\eta}$ is always within $\sim 2 \sigma_{\hat{\eta}}$ (shown by the grey bands) of $\eta$, regardless of the $\neff$. From \Cref{fig:eta-dist} we can see that $\neff=10^3$ leads to relatively small variability in both $\hat{\eta}$ and the bank size, while maintaining good computational efficiency in our tests throughout the sections below. We therefore recommend $\neff=10^3$ for $\eta\sim0.9$ (regardless of parameter space volume or dimensionality), although this must be adjusted according to the user's preferences.

Since $\eta$ is bounded from above ($\eta\leq1$), in principle the error should be asymmetric around $\hat{\eta}$. This is especially true for $\hat{\eta} \sim 1$ or for low numbers of effectualness points. The left hand panel of \Cref{fig:eta-dist} clearly shows the limitation of our error estimate where the grey $2\sigma_{\hat{\eta}}$ error band extends above $\eta=1$ when $\neff \sim \mathcal{O}(10)$. In practice however, we will always use $\neff = \mathcal{O}(10^3)$, where the error is well away from the boundary for $\eta=0.9$.

An additional impact of our stopping criterion is that the number of templates can vary between banks generated using the same waveform model and with the same input values for $\eta$ and $\targetmismatch$. As $\neff$ is increased, the variance in the number of templates required to meet the stopping criterion is reduced (illustrated in the right hand panel of \Cref{fig:eta-dist}). One therefore needs to choose a sufficiently large $\neff$ in order to minimize bank size variance but small enough to maintain fast generation (the default choices will be discussed below). See \Cref{sec:scaling} for a full description of the models used to generate \Cref{fig:eta-dist}.  

Finally, in order to actually sample from the probability distribution $p(\pintrinsic) \propto \sqrt{g(\pintrinsic)}$ associated with the metric, we employ rejection sampling~\cite{bda}. Rejection sampling requires selecting a proposal distribution $q(\pintrinsic)$ which is easy to sample from and for which there exists a constant $M$ such that $M \, q(\pintrinsic) \geq p(\pintrinsic)$ for all values of $\pintrinsic$. A sample from $p(\pintrinsic)$ can be generated using the following procedure:
\begin{itemize}
    \item Sample $\pextrinsic \sim q(\pintrinsic)$.
    \item Sample $u$ uniformly over the interval $[0, 1]$.
    \item If $u < p(\pintrinsic) / [M q(\pintrinsic)]$, return $\pintrinsic$. Otherwise, repeat the procedure.
\end{itemize}
We use a uniform distribution over the parameter space as the proposal. The constant $M$ is then equal to the maximum value of $\sqrt{g}$ over the parameter space. In practice, this point often lies on the boundary of the space and can be found numerically; alternatively, it can be estimated using empirical supremum rejection sampling~\cite{10.2307/4140534}. We note that for parameter spaces where the ratio between the square roots of the maximum and minimum values for the metric determinant is large, rejection sampling can become inefficient. Nevertheless, we have found that the metric evaluation is easily fast enough to quickly generate template banks in the parameter spaces typically considered for \gls*{gw}s. This sampling could be improved through importance sampling or using normalizing flows to learn the sampling distribution (see e.g.~Ref.~\cite{dinh_flows} and the recent review Ref.~\cite{flow_review}). We leave this to future work.

\subsection{Scaling and Coverage Properties}
\label{sec:scaling}

Here we study the scaling of the size and generation time for our random template banks. We begin by obtaining simple scaling relations to build rough intuition. Subsequently, in the remainder of this section, we will perform numerical experiments that illustrate how realistic banks deviate from these simple relations. We also compare the size and generation time of our banks with stochastic ones.

Let $\pcov$ denote the probability that a randomly-placed template covers a given effectualness point. For parameter spaces that are much larger than any template or with periodic boundary conditions, $\pcov$ is the ratio of the volume of a template to the volume of the space:
\begin{equation}
    \pcov = \frac{V_T}{V_S} = \frac{\targetmismatch^{n/2} V_n}{V_S} \, .
\end{equation}
The second equality expresses the template's volume in terms of the space's dimensionality $n$, the maximum mismatch $\targetmismatch$, and the volume of an $n$-dimensional unit sphere $V_n$.
For the waveforms we used in tests (described in the next section), we find this simple relation rarely holds due to boundary effects.
For example, if a template's ellipsoid extends significantly beyond the boundary of the parameter space in a given direction, the \emph{nominal} volume $V_T$ will differ from the \emph{actual} volume of parameter space the template covers~\cite{Manca:2009xw} (as illustrated in \Cref{fig:eff_points_illustration}).
In this case we do not expect $\pcov$ to scale precisely as $\targetmismatch^{n/2}$, and must instead use an \gls*{mc} approach to estimate $\pcov$.
Boundaries also make $\pcov$ position-dependent. In the most dramatic case we tested it can vary by nearly 100\% when the parameter space has narrow corners. 

Nevertheless, to obtain rough intuition, we first use a position-independent $\pcov$ to obtain simple scaling relations as follows. First we consider the expected size of our banks. Assuming the probability of covering multiple effectualness points with the same template is negligible, the average number of live effectualness points remaining after $N$ templates have been generated is $(1-\pcov)^N \, \neff$.\footnote{
    Note that this formula does not assume a given effectualness point is only covered by a single template.
} Bank generation terminates when the number of live points is less than or equal to $(1 - \eta) \, \neff$. Equating these quantities gives the average number of templates at termination:
\begin{equation}
    \Ntemplatesrandom = \frac{\log(1 - \eta)}{\log(1 - \pcov)} \, . \label{eq:n_templates-est}
\end{equation}
This is exactly the same scaling (as a function of $\eta$ and $\pcov$) as for random template banks described in Ref.~\cite{Messenger:2008ta}.

We can also estimate the average bank generation time. The cost of generating a template depends on the number of remaining live points, which is initially $\neff$. After generating a new template, the number of remaining live points is reduced by a factor of $1 - \pcov$ on average. This means the total computational cost is proportional to
\begin{align}
    C_\mathcal{R} \propto \sum_{k=1}^\Ntemplatesrandom (1 - \pcov)^{k-1} \, \neff = \neff \frac{1 - (1 - \pcov)^\Ntemplatesrandom}{\pcov} \, . \label{eq:time-random}
\end{align}

To make contact with the scaling properties of stochastic template banks, we modify our random bank procedure by adding a rejection step. This step requires comparing a proposal template with all the other templates in the bank and only adding it to the bank if its match with each of them is below $\operatorname{m}_*$. The other elements of our bank generation procedure remain the same (e.g. sampling templates according to the metric density and the convergence criterion). Note that this differs from the typical convergence criterion for stochastic banks, which involves waiting until the acceptance rate for new templates drops below a predetermined threshold (see e.g.~\cite{Harry:2009ea,Roy:2017oul,Roy:2017qgg}). Instead, we propose applying our convergence criterion to stochastic bank generation since it more directly connects with the coverage properties of the bank (i.e. $\eta$ and $\operatorname{m}_*$).

Generating a stochastic bank in this manner allows us to write a simple scaling relation for its generation time. Due to the rejection step, the average number of proposals required to generate the $k$th template is the inverse of the covering fraction, $(1 - \pcov)^{-(k-1)}$. Each of these proposals requires computing the match with the $k-1$ templates in the bank. The total bank generation cost is therefore
\begin{align}
    C_\mathcal{S} &\propto  C_\mathcal{R}|_{\Ntemplatesrandom \rightarrow \Ntemplatesstochastic}+ \sum_{k=1}^{\Ntemplatesstochastic} (k - 1) \, (1 - \pcov)^{-(k-1)} 
    \notag \\
    &=  C_\mathcal{R}|_{\Ntemplatesrandom \rightarrow \Ntemplatesstochastic} + \frac{(1 - \pcov)^{1 - \Ntemplatesstochastic} \left[ \Ntemplatesstochastic \, \pcov + (1 - \pcov)^{\Ntemplatesstochastic} - 1 \right]}{\pcov^2} \, , \label{eq:time-stochastic}
\end{align}
where $\Ntemplatesstochastic$ denotes the size of the stochastic bank and $\Ntemplatesrandom \rightarrow \Ntemplatesstochastic$ indicates that one needs to replace $\Ntemplatesrandom$ with $\Ntemplatesstochastic$ in \Cref{eq:time-random}. Unfortunately it is difficult to write a closed-form expression for $\Ntemplatesstochastic$, though it is always lower than $\Ntemplatesrandom$. Below we will therefore use $\Ntemplatesstochastic$ to denote the number of templates in a stochastic bank at termination, but note that this cannot be calculated a priori.

To experimentally check and compare these scaling relations, we use the \texttt{TaylorF2} waveform model with the spin contributions to the phase turned off. This yields a 3.5PN waveform in two dimensions which we parametrize using the black hole masses $m_1$ and $m_2$, with the restriction $m_1 > m_2$. We consider masses between \SI{1}{\Msun} and \SI{3}{\Msun}. For simplicity and speed we employ the analytic LIGO-I noise power spectral density from table~IV of Ref.~\cite{Damour:2000zb}, defined for $f > f_s$ as
\begin{align}
    S_n&(f) = \SI{e-46}{\hertz^{-1}} \notag \\
    & \times 9 \left[ (4.49x)^{-56} + 0.16 x^{-4.52} + 0.52 + 0.32 x^2 \right] \, ,
\end{align}
where $x \equiv f / f_0$, $f_0 = \SI{150}{\hertz}$, and the lower cutoff is $f_s = \SI{40}{\hertz}$. We use a frequency range of \SIrange{40}{512}{\hertz} and spacing of $\Delta f = \SI{0.1}{\hertz}$. Finally, we use $\neff = 1000$ for all tests. To estimate the covering probability $\pcov$, we first randomly sample \num{10000} independent pairs of templates and points according to the metric density. For each pair we then check whether the template covers the corresponding point and take the total fraction of covered points to be our estimate of $\pcov$. Over the range $\operatorname{m}_* \in [0.75, 0.95]$, this \gls*{mc} estimate yields values between $\pcov = 0.01$ and $\pcov = 0.0023$. 

\begin{figure*}
    \centering
    \includegraphics[width=0.85\textwidth]{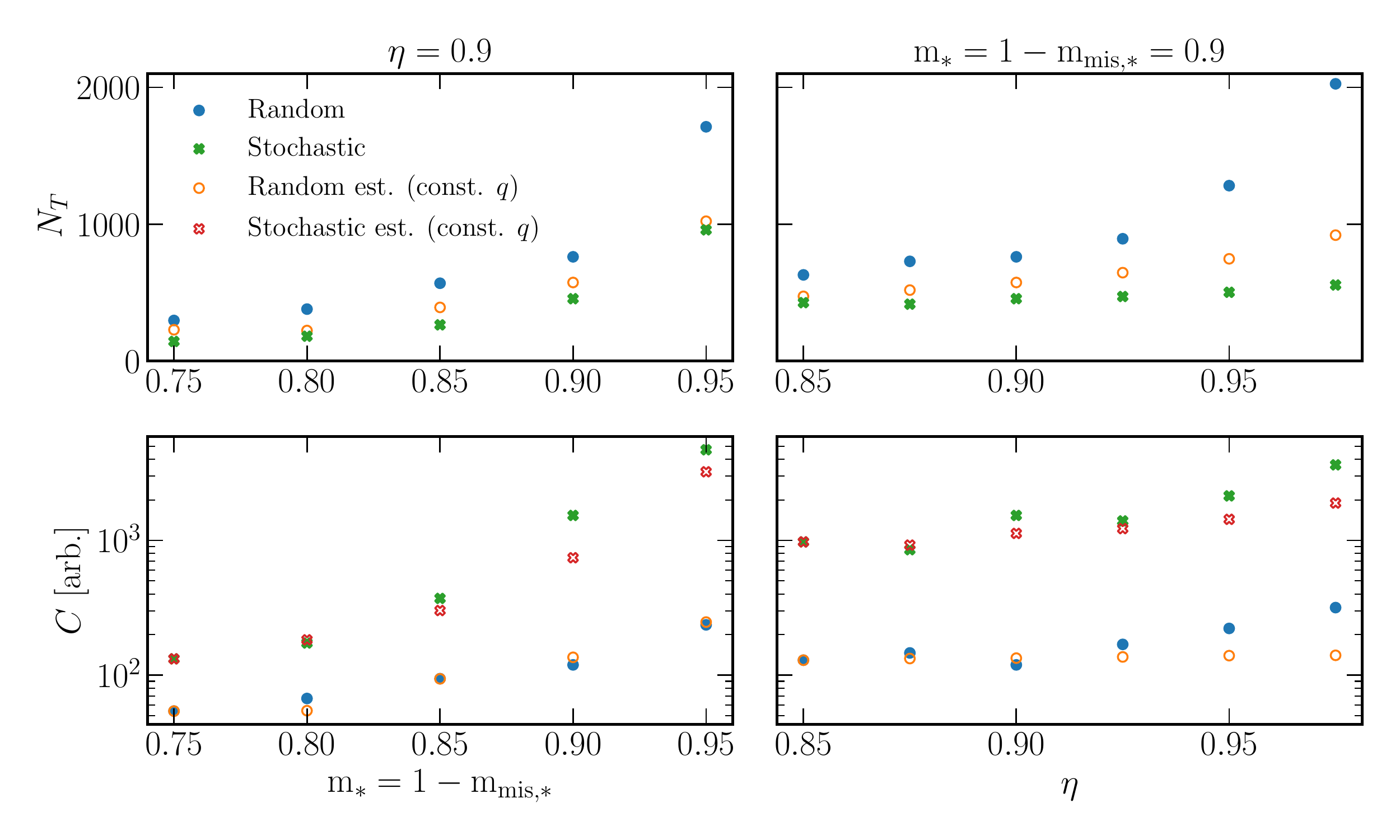}
    \caption{Scaling of bank size and cost of generation as a function of minimum match and target covering fraction. The analytic estimates (orange points) come from \Cref{eq:n_templates-est} (top row) and \Cref{eq:time-random,eq:time-stochastic} (bottom row). Note that our analytic estimates start to differ from our real banks for $(1-\targetmismatch)\rightarrow 1$ and $\eta\rightarrow 1$. This is due to the growing importance of a spatially dependent $\pcov$ (see text for more details). In the bottom row, one can see that the cost of generating a stochastic bank is significantly greater than a random bank. This difference continues to grow with larger template banks. All plots are based on the two-dimensional 3.5PN model and analytic model for the noise power spectral density described in the text. \href{https://github.com/adam-coogan/diffbank/blob/paper-1/scripts/plot_scaling.py}{\faFileCodeO}}
    \label{fig:scaling}
\end{figure*}

The number of templates and cost of generation for both stochastic and random banks are plotted in \Cref{fig:scaling}. We also plot our estimates for these quantities from \Cref{eq:n_templates-est} (top row) and \Cref{eq:time-random,eq:time-stochastic} (bottom row). In applying \Cref{eq:time-random,eq:time-stochastic} we fix $\Ntemplatesrandom$ to the true sizes of the template banks rather than using the estimate from \Cref{eq:n_templates-est}. For most choices of the covering fraction and maximum mismatch, the sizes of the random and stochastic banks differ by less than a factor of two. However, for large values of the covering fraction and small values of the maximum mismatch, this can increase to a factor of four. The CPU time required to perform a search with one of our template banks would thus be correspondingly larger than with a stochastic bank. On the other hand, the generation time of the random banks is a factor of a few to over an order of magnitude faster than for the stochastic banks.

Our scaling relations approximately hold for high maximum mismatches and low covering fractions. However, they deviate from the experimental results at higher values of those parameters (i.e. when $(1-\targetmismatch)\rightarrow 1$ or $\eta\rightarrow 1$). The reason is that our relations ignore the fact that the covering probability $\pcov$ varies dramatically over the parameter space due to boundary effects. More precisely, the probability of covering the final few effectualness points is substantially lower than the averaged value of $\pcov$ we use to derive the estimates in the figure, driving up the bank size and generation time. This is particularly true for high $\eta$ since a higher fraction of the total points need to be covered, therefore emphasizing the regions with a lower $\pcov$ (for fixed $\targetmismatch)$. On the other hand, lower values of $\targetmismatch$ (with fixed $\eta$) lead to an increased variation of $\pcov$ across the parameter space which in turn leads to a similar inaccuracy in scaling estimates.

It is worth noting that the probability of accepting a template during stochastic bank generation can vary significantly depending on the proposal distribution used.\footnote{Here we define the proposal distribution as the distribution used to generate proposal points (which are then either accepted or rejected) during stochastic bank construction.} In particular, if the proposal distribution significantly differs from the metric density, the average number of proposals needed to accept a template will increase~\cite{Manca:2009xw} (i.e., $\pcov$ will decrease) and therefore the associated cost of generating a stochastic bank will also increase. Since the metric evaluation using \gls*{ad} is so cheap, we therefore recommend using the metric density to improve the efficiency of stochastic bank generation.

\subsection{Mean Mismatch}

A key quantifier for a template bank is the maximum mismatch between a waveform which lies within the range of parameters covered by the bank, and the bank itself. 
Geometric template banks can enforce a maximum mismatch, which in turn describes the maximum loss of \gls*{snr} that can occur due to the discreteness of the bank when carrying out matched-filtering searches.
For stochastic and random template banks, no guarantee is made for the maximum mismatch, and instead the appropriate quantity is the expected mismatch $\langle \mismatch \rangle$, as a function of $\eta$ and $\targetmismatch$. For random template banks, a closed form expression can be derived when boundaries are neglected, using the probability density for the mismatch given in Ref.~\cite{Messenger:2008ta}. The expected mismatch is
\begin{align}
    \langle \mismatch \rangle = \frac{\targetmismatch}{2} &\sqrt{\frac{\pi}{-\log(1 - \eta)}}\erf\left( \frac{\sqrt{-\ln(1 - \eta)}}{\targetmismatch} \right) \notag \\
    & - (1-\eta)^{1/\targetmismatch^2} \, .
    \label{eqn:expectedm}
\end{align}
This expression can be used as a rough guide to select the parameters required for a desired $\langle \mismatch \rangle$.

Although \Cref{eqn:expectedm} is derived neglecting boundary effects, it should approximately apply when the total template volume outside of the boundaries of the bank is smaller than the volume contained within the boundaries. In the case where boundary effects are important (i.e.~a substantial fraction of the total template volume extends out of bounds), this expression will overestimate the mean mismatch. This is because the portion of the template ellipsoid that lies out of bounds has a higher mismatch. The regions remaining in bounds will thus have a lower mean mismatch, making the analytic expression an overestimate of the true mean mismatch.

\section{Comparisons with Existing Methods and Banks}
\label{sec:comparison}

In this section we generate template banks for real waveforms in two, three, and four dimensional parameter spaces and compare to existing template banks in the literature. We additionally demonstrate that search-ready random banks can be generated with little computational overhead using \diffbank~\cite{diffbank}.

\subsection{Realistic Banks}

Here we consider three waveforms with varying degrees of complexity and dimensionality. To easily discuss the different waveforms, we name each waveform according to its \gls*{pn} order and dimensionality of parameter space. Below we provide the waveform names as well as a short description of the waveform and parameter space:
\begin{itemize}
\itemsep0em
    \item 3.5PN-2D --- First, we consider the \texttt{TaylorF2} waveform model but ignore all contributions from spin to the phase of the waveform. In particular, it is a 3.5PN waveform in two dimensions (we use $m_1$ and $m_2$ for the mass of each binary object and enforce $m_1 > m_2$). Here we consider $\SI{1}{\Msun} \leq m_1, \, m_2 \leq \SI{3}{\Msun}$. This is the same waveform employed in the previous section to study the scaling and coverage properties of random banks. We were unable to find a suitable bank from the literature to compare with; we therefore construct our own stochastic bank with the same $\targetmismatch$ and $\eta$.
    \item 2.5PN-3D --- Second, we look at a 2.5PN waveform in three dimensions introduced in Ref.~\cite{Ajith:2012mn}. Importantly, this waveform model adds an additional spin parameter $\theta_{3S}$ which accounts for aligned spin components for both objects. While the waveform is parametrized in terms of the dimensionless chirp times $(\theta_0, \theta_3)$ and this spin parameter $\theta_{3S}$, the boundaries are defined in terms of the physical properties of the components of the system. As explained in Table~I of Ref.~\cite{Ajith:2012mn}, the component masses are restricted to the interval $[1,\, 20]\, \Msun$ and the total mass is fixed between $[2,\, 21]\, \Msun$. Objects with mass below $\SI{2}{\Msun}$ are considered neutron stars, with spin parameters $\chi$ restricted to the range $[-0.4,\, 0.4]$. Heavier objects are considered black holes with spin parameters restricted to $[-0.98,\, 0.98]$. The cutoff frequency $f_0$ used to define the chirp times is set to \SI{20}{\hertz}.
    \item 3.5PN-4D --- Finally, we consider the \texttt{TaylorF2} waveform model which, in addition to the black hole masses, has parameters describing the magnitude of the aligned spin components of each black hole, $\chi_{1,2}$. This model is typically used to search for low-mass binary signals in current LIGO and Virgo analyses (see e.g.~\cite{LIGOScientific:2020ibl,LIGOScientific:2021djp} and the references therein) and therefore represents our current state-of-the-art. We use the same mass ranges as for 3.5PN-2D and additionally consider $-0.99 \leq \chi_{1,2} \leq 0.99$. These ranges were chosen to directly compare to the binary neutron star banks generated in Ref.~\cite{Roulet:2019hzy}.
\end{itemize}
For the 2D and 4D banks we use the frequency range \SIrange{24}{512}{\hertz}, $\Delta f = \SI{0.1}{Hz}$, and $\neff = 1000$. For the 3D bank we instead use the frequency range \SIrange{20}{2200}{\hertz} with the same frequency spacing (for consistency with the reference bank) and $\neff = 1300$ (due to the large value of $\eta$ used by the reference bank). The noise models, $\eta$ values, and $\targetmismatch$ values for each waveform are listed in \Cref{tab:banks}. These were chosen to align as closely as possible to those used to generate the banks with which we are comparing.

In \Cref{tab:banks} we show the resulting generation times and bank sizes. The banks were made using a single NVidia V100SXM2 graphical processing unit with 16GB of memory using 64-bit floating point precision. For comparison we also list the sizes of reference banks generated with similar values of $\targetmismatch$ and $\eta$.\footnote{
    $\eta$ is not an explicit parameter used to generate the 2.5PN-3D and 3.5PN-4D reference banks. Instead we extract these from the caption of Fig.~3 in Ref.~\cite{Ajith:2012mn} and the binary neutron star curves in the upper panel of Fig.~5 in Ref.~\cite{Roulet:2019hzy} respectively.
} For the 3.5PN-2D model, the reference bank is a stochastic bank we generated ourselves. The reference banks for the 2.5PN-3D and 3.5PN-4D models are the stochastic bank from Ref.~\cite{Ajith:2012mn} and the ``geometric placement'' bank from Ref.~\cite{Roulet:2019hzy}.

For all three waveforms, to achieve the same covering fraction $\eta$,  the number of templates required in the reference banks is smaller than the number in our random banks. The stochastic 3.5PN-2D (reference) bank contains about 30\% fewer templates than the corresponding random bank. Our other random banks' sizes are within a factor of 3.8 of the more minimal reference ones. We also note that the higher value of $\eta$ used for the 2.5PN-3D model caused the random template bank to be significantly larger than the reference stochastic bank. This feature is expected based of the trend seen in \Cref{fig:scaling}, and illustrates that random banks grow quickly for $\eta \rightarrow 1$. Since smaller bank sizes are preferable, it is therefore advisable to use a sufficiently high $\eta$ to cover a significant fraction of the parameter space, but not so high that the random bank is too large.  Based off of the scaling in \Cref{fig:scaling}, we see that the bank sizes start to deviate significantly at $\eta \sim 0.9$ and therefore advise using a similar value.

From the above discussion, it is clear that  random banks are \emph{not} optimal: they do not use the minimum number of templates to cover the maximum amount of parameter space. However, our random banks have several major advantages.  They are much more efficient to generate than the reference banks. For the 3.5PN-2D stochastic reference bank, the generation time was over 250 times longer (\SI{33}{\hour} \SI{45}{\minute} \SI{45}{\second}) than for our random bank on the same hardware.  Indeed our random banks are extremely simple to generate in practice and for higher dimensions may even become significantly more efficient than optimal lattice placement schemes~\cite{Messenger:2008ta}. In the future, it is expected that more sophisticated waveforms of ever-higher dimensionality (more parameters) will be used in the analysis of LIGO and Virgo data. Hence our method of template bank generation will become ever more important due to its efficiency.

\begin{figure}
    \centering
    \includegraphics[width=\linewidth]{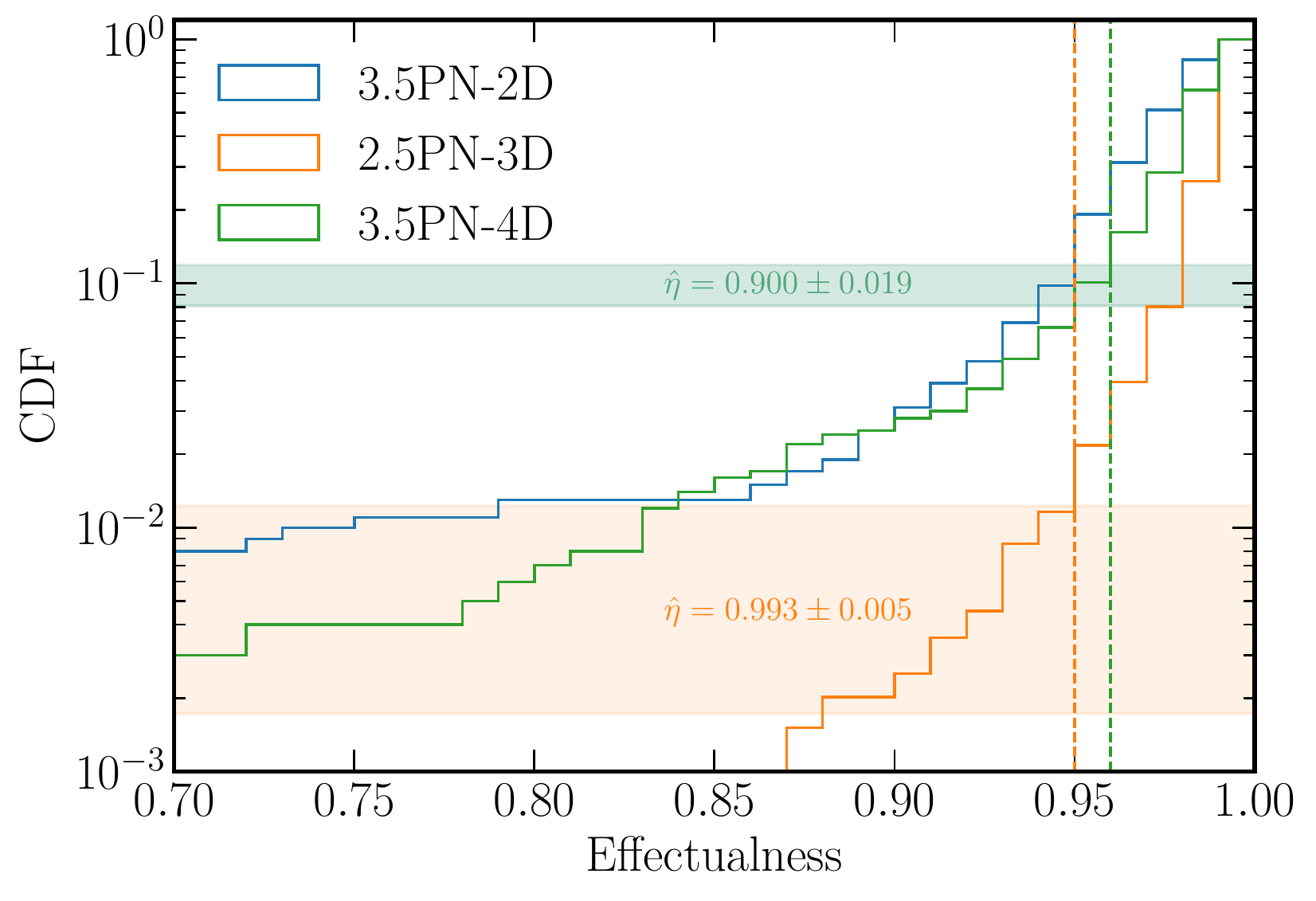}
    \caption{Cumulative distribution function for effectualnesses of each of the three different template banks as labeled in the legend. In each case the effectualnesses were computed at \num{1000} points sampled from the metric density. The vertical lines show the target values of $1-\targetmismatch$ for each bank; note that the blue and orange lines overlap. The bands show the CDF corresponding to the target value of $\eta$ plus and minus $2\sigma_{\hat{\eta}}$, \Cref{eq:sigma-eta-hat}; the blue and green bands overlap. \href{https://github.com/adam-coogan/diffbank/blob/paper-1/scripts/plot_bank_effs.py}{\faFileCodeO}}
    \label{fig:bank-effs}
\end{figure}

Finally, in \Cref{fig:bank-effs} we show the cumulative distribution function of the effectualness calculated for $1000$ randomly sampled points in the parameter space for all three banks. The dashed vertical lines indicate the banks' values of $\targetmismatch$. The horizontal bands show $1 - (\hat{\eta} \pm 2 \sigma_{\hat{\eta}})$ -- i.e., the CDF corresponding to the target covering fraction with uncertainties coming from using a finite number of effectualness points. All our banks achieve a covering fraction within this error band at their values of $\targetmismatch$. This can be seen from the plot as the CDFs for each bank pass through the corresponding vertical line within the corresponding band.

\begin{table*}
    \centering
    \begin{tabular}{c c c c c c c c}
        \toprule
         Name & $\operatorname{m}_*$ & $\eta$ & Frequency range & Noise model & $T_{\rm gen}$ & $\Ntemplatesrandom$ & $N_T^\mathrm{ref}$ \\
         \midrule
         3.5PN-2D & 0.95 & 0.9 & \SIrange{24}{512}{\hertz} & LIGO Livingston O3a\footnote{\url{https://dcc.ligo.org/LIGO-P2000251/public}} & \SI{7}{\minute} \SI{55}{\second} & \num{10780} & \num{7197} \\
         2.5PN-3D & 0.95 & 0.993 & \SIrange{20}{2200}{\hertz} & \texttt{aLIGOZeroDetHighPower}\footnote{From \texttt{pycbc}~\cite{pycbc}} & \SI{64}{\hour} \SI{17}{\minute} \SI{44}{\second} & \num{2075173} & \num{549194}~\cite{Ajith:2012mn} \\
         3.5PN-4D & 0.96 & 0.9 & \SIrange{24}{512}{\hertz} & LIGO O2\footnote{\url{https://github.com/jroulet/template_bank/}} & \SI{51}{\minute} \SI{46}{\second} & \num{280967} & \num{116443}~\cite{Roulet:2019hzy} \\
         \bottomrule
    \end{tabular}
    \caption{Results of our random template bank generation tests for three waveforms described in the text. The first column is the name of the bank in terms of PN order and dimensionality of parameter space. The quantities $\operatorname{m}_*$ and $\eta$ are the chosen values of the target minimum match and target covering fraction of the template bank. For the given frequency range and noise model, $T_{\rm{gen}}$ is the generation time for our random template bank, and $\Ntemplatesrandom$ is the number of templates in our bank. For comparison, the last column indicates the size $N_T^\mathrm{ref}$ of reference banks. For the 3.5PN-2D model the reference bank is a stochastic bank that we ourselves generated; for the other models, the reference bank size is taken from the literature as indicated. All our banks were generated on an NVidia V100SXM2 with 16GB of memory using 64-bit floats.}
    \label{tab:banks}
\end{table*}

\section{Discussion and Conclusion}
\label{sec:conclude}

Data from existing and planned gravitational wave detectors promise to be a goldmine for refining our understanding of astrophysics, astronomy, and fundamental physics. This work addresses the problem of constructing template banks for generic frequency domain waveforms, enabling searches for new types of compact binary coalescence signals using matched filtering pipelines.  To date, matched-filter searches have focused on aligned-spin \gls*{bbh} systems on quasi-circular orbits. Our goal is to allow more general searches that might potentially lead to discovery of new astrophysics and physics Beyond the Standard Model.

In matched filtering, the strain data are compared to a bank of templates described by a set of points in the binary parameter space and a \gls*{gw} waveform model.
The foundation of our new bank generation scheme is differentiable waveforms, which make it possible to efficiently compute the parameter space metric for gravitational waveforms using automatic differentiation. We use this metric to implement a new variant of random template bank generation~\cite{Messenger:2008ta} that uses a set of fixed effectualness points to monitor the bank's coverage properties. A schematic illustration of our template bank generation procedure is shown in Figure 3. Starting from $\neff$  effectualness points sampled according to the metric density, we sequentially add templates until a predetermined fraction $\eta$ of these points is covered, i.e. the match between the point and a given template exceeds a minimum value $1 - \targetmismatch$ (typically we chose $\eta \sim 0.9$ and $1-\targetmismatch \sim 0.95$).  

This approach has several advantages:
\begin{itemize}
\itemsep0em
    \item Computing the metric with automatic differentiation removes the need to derive the metric by hand or by use of numerical differentiation, which can be noisy and involve many waveform evaluations.
    \item The generation time for our banks scales much more favorably than for stochastic banks as a function of the parameters controlling its effectualness coverage ($\targetmismatch$ and $\eta$). This was borne out by our numerical experiments. In combination with our use of the \jax automatic differentiation framework, this enables rapid generation of template banks using CPUs or GPUs.
    \item Our new approach to monitoring convergence removes the precalculation of bank size previously required for random bank generation~\cite{Messenger:2008ta} which is challenging to perform for waveforms where the template volume extends well beyond the parameter space boundaries (c.f. \Cref{sec:generating-a-bank}). Our method additionally provides a Monte Carlo error estimate for the fraction of the parameter space covered by the bank.
    \item Finally, we go beyond the random banks studied in Ref.~\cite{Messenger:2008ta} by accounting for parameter space boundaries. In particular, using effectualness points to calculate the coverage of the template bank naturally accounts for parameter space boundaries without any fine tuning. This feature is essential for using random banks for realistic \gls*{gw} waveforms.
\end{itemize}

We also compared with template banks from the literature that use realistic waveform models of different dimensionalities and with different detector noise models. We found that we could rapidly generate comparable random banks on a single GPU. We also showed that $\neff = 1000$ is sufficient to accurately monitor covering fraction of the bank such that the realized covering fraction $\hat{\eta}$ is close to the target value $\eta$.

Our approach inherits some of the advantages and disadvantages of other random template bank methods. In comparison to lattice banks our random banks do not fully cover the parameter space, but are much simpler to generate since they work in curved parameter spaces with arbitrary boundaries. In comparison with the more widely-used stochastic banks, our random banks are larger by a factor of $\sim 1.5$ to $\sim 3.75$ in our experiments. While this correspondingly increases the \emph{CPU time} required for searches using our banks, this may not result in a correspondingly larger \emph{wall time}, since matched filtering searches are straightforward to parallelize. On the other hand, our banks are much faster to generate than stochastic banks. Since bank generation is more difficult to parallelize, this helps counteract the search wall time increase caused by our larger banks.

To make random banks more efficient, one could implement a secondary \emph{pruning} step which removes unnecessary templates. Unfortunately, all such pruning calculations are likely to require a significant number of match calculations and may be as computationally expensive as simply constructing a stochastic bank from the beginning.\footnote{
    One could also imagine optimizing the final random bank in order to increase $\eta$~\cite{Fehrmann:2014cpa}, although this may be even more computationally expensive than a pruning step.
} We leave a more detailed investigation of a pruning step to future work.

The use of automatic differentiation to calculate the metric comes with some practical restrictions on the form of the waveform, discussed in \Cref{sec:limAD}. We expect these restrictions will loosen as \jax and other automatic differentiation frameworks mature. Another potential issue is the strong hierarchy of parameters (i.e., the chirp mass is the main parameter governing the shape of the waveform while the mass ratio has little impact), which can cause the metric to be poorly conditioned, leading to instabilities in the calculation of its determinant. We expect this could be alleviated through automatically learning new waveform parametrizations (akin to chirp times), though we leave this for future work (see also Ref.~\cite{Roy:2017qgg}).

Lastly, we have implemented our template bank generator in the easy-to-use \diffbank package to enable physicists to rapidly create template banks for their favorite waveform models. We are currently utilizing this tool to construct a template bank to search for objects with enhanced spin-induced quadrupoles in LIGO data~\cite{waveformpaper,searchpaper}. We hope that \diffbank spurs the community to perform searches for other novel compact objects in the new world of gravitational wave data and to investigate other uses for differentiable waveforms.

\begin{acknowledgments}
    We thank Chris Messenger for useful discussions.

    A.C.~acknowledges support from the Schmidt Futures foundation. A.C.~and C.W.~received funding from the Netherlands eScience Center, grant number ETEC.2019.018. T.E.~and K.F.~acknowledge support by the Vetenskapsr{\aa}det (Swedish Research Council) through contract No.~638-2013-8993 and the Oskar Klein Centre for Cosmoparticle Physics. H.S.C.~gratefully acknowledges support from the Rubicon Fellowship awarded by the Netherlands Organisation for Scientific Research (NWO). K.F.~is Jeff \& Gail Kodosky Endowed Chair in Physics at the University of Texas at Austin, and is grateful for support. K.F.~acknowledges funding from the U.S. Department of Energy, Office of Science, Office of High Energy Physics program under Award Number DE-SC0022021 at the University of Texas, Austin. C.M.~and A.Z.~were supported by NSF Grant Number PHY-1912578. C.W.~received funding from the European Research Council (ERC) under the European Union’s Horizon 2020 research and innovation program (Grant agreement No.~864035).
    
    This material is based upon work supported by NSF's LIGO Laboratory which is a major facility fully funded by the National Science Foundation.
    
    This research was enabled in part by support provided by Calcul Québec (\url{https://www.calculquebec.ca/}) and the Digital Research Alliance of Canada (\url{https://alliancecan.ca/}). The Béluga cluster on which the computations were carried out is 100\% hydro-powered.
    
    Finally, we acknowledge the use of the Python modules \jax~\cite{jax2018github}, \texttt{jupyter}~\cite{jupyter}, \texttt{matplotlib}~\cite{Hunter:2007}, \texttt{numpy}~\cite{numpy}, \texttt{scipy}~\cite{scipy} and \texttt{tqdm}~\cite{tqdm}.
\end{acknowledgments}

\bibliography{main.bib}

\appendix

\section{Monte Carlo error estimation for the covering fraction \texorpdfstring{$\eta$}{eta}}
\label{sec:eta-error}

In this appendix we explain how to derive the Monte Carlo error on our estimate of $\hat{\eta}$ (\Cref{eq:sigma-eta-hat}).

Define the function $c(\pintrinsic)$ over the parameter space as being equal to one if $\pintrinsic$ is covered by a template in a bank and equal to zero otherwise. The \emph{true covering fraction} $\eta$ of the bank is obtained by averaging $c$ over the whole parameter space (i.e., integrating and dividing by the space's volume $V$). This can also be approximated through \gls*{mc} integration by randomly sampling a set $\Theta$ of $n$ parameter points uniformly over the space and averaging $c$ over them, yielding $\hat{\eta}$:
\begin{align}
    \eta = \frac{1}{V} \int \dd{\pintrinsic} c(\pintrinsic) \approx \frac{1}{\neff} \sum_{\pintrinsic \in \Theta} c(\pintrinsic) = \hat{\eta} \, .
\end{align}

\gls*{mc} error estimates are typically obtained using the \gls*{clt}. By the \gls*{clt}, were we to repeatedly compute $\hat{\eta}$ with large enough $n$, the values would follow a normal distribution $\mathcal{N}(\eta, \sigma^2 / n)$, where $\sigma^2$ is the variance of $c$:
\begin{equation}
\begin{split}
    \sigma^2 &= \frac{1}{V} \int \dd{\pintrinsic} \left[ c(\pintrinsic) - \eta \right]^2 \\
    &= \frac{1}{V} \int \dd{\pintrinsic} \left[ c(\pintrinsic)^2 - 2 c(\pintrinsic) \eta + \eta^2 \right] \\
    &= \frac{1}{V} \int \dd{\pintrinsic} c(\pintrinsic) - \eta^2 \\
    &= \eta (1 - \eta) \, ,
\end{split}
\end{equation}
where we used the fact $c^2(\pintrinsic) = c(\pintrinsic)$. This implies that $\sigma_{\hat{\eta}} \equiv \sigma / \sqrt{n} = \sqrt{\eta (1 - \eta) / n}$ estimates the error on $\hat{\eta}$. Since the true values of $\sigma$ and $\eta$ are unknown, we instead approximate them with the mean and standard deviation of $c$ evaluated over our set of points $\Theta$.

Since our bank generation procedure gives a \gls*{mc} estimate of $\eta$ by tracking how many effectualness points have been covered, we can also use the logic above to determine the accuracy of this estimate. While the conditions of the \gls*{clt} are not strictly satisfied since values of $c$ evaluated at each effectualness point are not independent due to the stopping criterion, we apply it regardless. Since bank generation stops when a fraction $\eta$ of effectualness points are covered, we have $\hat{\eta} = \ceil{\eta \neff} / \neff \approx \eta$. This implies the set of covered points $\Theta_1$ has size $\ceil{\eta \neff}$ and the set of uncovered points $\Theta_0$ has size $\neff - \ceil{\eta \neff}$. Then our estimate for the error on $\hat{\eta}$ is
\begin{equation}
\begin{split}
    \sigma_{\hat{\eta}} &= \sqrt{\frac{1}{\neff} \frac{1}{\neff - 1} \sum_{\pintrinsic \in \Theta} \left[ c(\pintrinsic) - \eta \right]^2} \\
    &= \sqrt{\frac{1}{\neff} \frac{1}{\neff - 1} \left[ \sum_{\pintrinsic \in \Theta_0} \eta^2 + \sum_{\pintrinsic \in \Theta_1} (1 - \eta)^2 \right]} \\
    &= \sqrt{\frac{1}{\neff} \frac{(\neff - \ceil{\eta \neff}) \eta^2 + \ceil{\eta \neff} (1 - \eta)^2}{\neff - 1}} \\
    &\approx \sqrt{\frac{\eta (1 - \eta)}{\neff - 1}} \, ,
\end{split}
\end{equation}
where we assumed $\eta$ was large enough that $\ceil{\eta \neff} \approx \eta \neff$, and the $\neff - 1$ factor is Bessel's correction for estimating the variance from samples. This is the equation we sought to derive (\Cref{eq:sigma-eta-hat}).

\end{document}